\documentclass[12pt,preprint]{aastex}
\begin{document}
\title{Cluster-Supercluster Alignments} 
\author{\sc Jounghun Lee\altaffilmark{1} and 
August E. Evrard\altaffilmark{2,3,4}}
\altaffiltext{1}{Department of Physics and Astronomy, FPRD, Seoul National 
University, Seoul 151-747 , Korea}
\altaffiltext{2}{Department of Astronomy and Michigan Center for Theoretical 
Physics, University of Michigan, Ann Arbor, 500 Church St., MI 48109, USA}
\altaffiltext{3}{Department of Physics and Michigan Center for Theoretical 
Physics, University of Michigan, Ann Arbor, 450 Church St.,MI 48109-1040, USA}
\altaffiltext{4}{Visiting Miller Professor, Physics Department, 
University of California, Berkeley, CA94720, USA}
\begin{abstract} 
We study correlations in spatial orientation between galaxy clusters and 
their host superclusters using a Hubble Volume N-body realization of a 
concordance cosmology and an analytic model for tidally-induced alignments.  
We derive an analytic form for distributions of the alignment angle as 
functions of halo mass ($M$), ellipticity ($\epsilon$), distance ($r$) and 
velocity ($v$) and show that the model, after tuning of three parameters, 
provides a good fit to the numerical results.  The parameters indicate a 
high degree of alignment along anisotropic, collapsed filaments.  
The degree of alignment increases with $M$ and $\epsilon$ while it 
decreases with $r$ and is independent of $v$.  We note the possibility of 
using the cluster-supercluster alignment effect as a cosmological probe to 
constrain the slope of the initial power spectrum.
\end{abstract} 

\section{INTRODUCTION}

Superclusters are collections of galaxy groups and clusters that
represent the largest, gravitationally bound structures in the universe
\citep{sha30,kal-etal98}.  If the dark energy is a cosmological
constant, then the collapse of these systems over the next few billion
years of the cosmic future will mark the end of hierarchical structure 
formation in our universe \citep{nag-loeb03,busha05}. A conspicuous feature 
of locally observed superclusters is the strong tendency of member clusters 
to be elongated in their major axis orientations \citep{pli02,pli04}, 
which is in turn closely related to their filamentary shapes 
\citep[e.g.,][]{bas03}. To describe the structure distribution
on the largest scale in the universe,  it will be quite essential to
understand this effect of cluster-supercluster alignments from
first principles.

The effect of structure-substructure alignment is in fact 
observed on all different scales in the universe. On the subgalactic
scale the galaxy satellites are observed to be located preferentially
near the major axes of their host galaxies
\citep{val-etal78,kne-etal04,bra05,agu-bra06}. On the galactic scale 
the major axes of cluster galaxies are observed to be aligned with
that of their host clusters \citep{pli-bas02,pli-etal03}. The cluster 
galaxies are also observed to have a strong tendency toward radial 
alignment \citep{per-kuh05}. 

Although this alignment effect has been shown to be a natural outcome
in the currently favored concordance $\Lambda$CDM cosmology
\citep{onu-tho00,lib-etal05,kan-etal05,lee-etal05,zen-etal05,kas-evr05,
bas-etal06}, its detailed origin remains a subject of debate between those 
who emphasize the importance of anisotropic merging and those who stress 
tidal interaction. 

The anisotropic merging scenario explains that the effect of
substructure-structure alignment is induced by the anisotropic
merging and infall of matter along filaments \citep{wes89}.
It was indeed shown by N-body simulations that the merging and infall 
of matter to form bound halos indeed occur preferentially along filaments, 
which provided supporting evidences for this scenario
\citep[e.g.,][]{wes-etal91,van-van93,dub98,fal-etal02,kne-etal04,zen-etal05}.

The tidal interaction theory explains that the correlations between the 
substructure angular momentum vectors and the principal axes of the 
host tidal fields induce the alignment effect. \citet{lee-etal05} constructed 
an analytic model for the effect of substructure alignment in the frame of the 
tidal interaction theory, and showed that their analytic predictions are
in good agreement with the numerical results from N-body simulations. 

In fact, the above two theories are not mutually exclusive since the 
anisotropic merging and infall itself is a manifestation of the primordial 
tidal field \citep{bon-etal96}. The difference between the two scenarios, 
however, lies in the question of whether the connection to filaments is a 
major contribution or not.

Very recently, \citet{atl-etal06} have quantified the influences 
of both the tidal interaction and the anisotropic infall through the
analysis of data from recent high-resolution N-body simulations. What they 
confirmed is the following: (i) For the majority of halos the alignment effect 
is caused by the tidal field but not by the anisotropic infall; (ii) Only for 
the cluster-size halos the alignment effect is dominantly due to the 
anisotropic merging and infall of matter along filaments. In other words, 
they made it clear that the filaments are important marker of local 
orientations on the cluster halo scale.

Now that the cluster-supercluster alignment turns out to be due to the
anisotropic merging along filaments, it is desirable to have a
theoretical frame work within which one can provide physical answers to the
remaining questions such as how the alignment effect depends on the
cluster properties such as mass, shape, and etc.  Our goal here is to
construct such a theoretical framework by using both analytical
and numerical methods. Analytically we adopt the standard  
cosmic web theory, and numerically we use the large Hubble volume
simulation data.

The organization of this paper is as follows. In $\S 2$ we provide a
brief description of the Hubble volume simulation and summarize the
numerical results. In $\S 3$ we present an analytic model and compare
its predictions with the numerical results. In $\S 4$ we discuss our
results and draw final conclusions.

\section{NUMERICAL RESULTS}
 
For the numerical analysis, we use a mass-limited sample of cluster
halos extracted from the Hubble Volume simulation of a $\Lambda$CDM
universe \citep{evr-etal02}.  The simulation models dark matter structure 
resolved by particles of mass $m=2.25\times 10^{12}h^{-1}M_{\odot}$ in a 
periodic cube of linear size $3000h^{-1}$Mpc, assuming 
$\Omega_m = 0.3,\Omega_{\Lambda}=0.7$ and $\sigma_{8}=0.9$. 
The $z=0$ catalog contains a total of $82973$ halos with 
mass above a limiting value $M > 3\times 10^{14}h^{-1}M_{\odot}$, 
with information on various properties such as center-of-mass position, 
mass, inertia momentum tensor, and redshift. We refer the readers to 
\citet{evr-etal02} and \citet{kas-evr05} for the details of the cluster 
catalog, including the algorithm of cluster identification. 

The superclusters are identified in the catalog with the help of the 
friends-of-friends algorithm with the linking length of $0.33\bar{l}$,
where $\bar{l} = 69h^{-1}$Mpc is the mean spacing of the mass-limited
sample. The total number and the mean mass of the identified superclusters 
are $N_{s}=14007$ and $\bar{M}_{s}=1.26 \times 10^{15}h^{-1}M_{\odot}$, 
respectively. This large number of superclusters allows us to study 
the alignment effect with high statistical power.

\begin{figure}
\begin{center}
\plotone{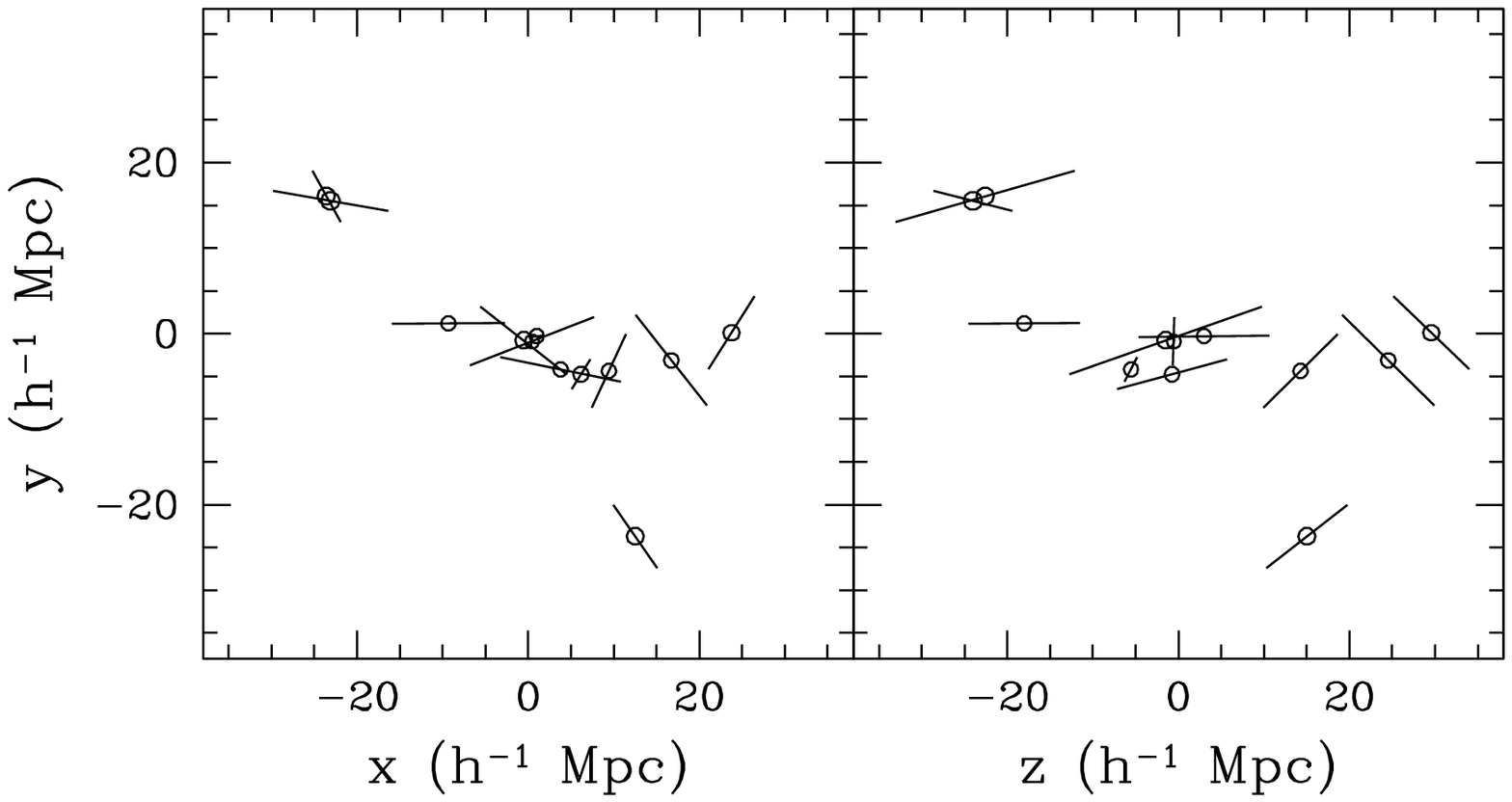}
\caption{The spatial distribution of the third richest supercluster is
  shown in orthogonal projections.  Circles show halo locations, with
  symbol size scaling as $M^{1/3}$, while lines through each halo show the
  orientation of the major axis of its density field, 
  taken from \citep{kas-evr05}.  The length of each line is
  proportional to the halo's major-to-minor axis ratio.
\label{fig:sc3}}
\end{center}
\end{figure}

Figure~\ref{fig:sc3} shows orthogonal projections of the third
richest supercluster in the volume.  It contains 12 halos above the 
applied mass limit, and a total mass of $5.3 \times 10^{15}
h^{-1} \, {\rm M}_{\odot}$ associated with these halos.  The spatial
distribution of the supercluster is highly elongated, much closer to
filamentary than spherical.  In this example, the major axis orientations 
of the halos, taken from \cite{kas-evr05}, are shown as whiskers in the plot. 
The tendency for these halos to be aligned with their supercluster's principal 
axis, although arguably visible in this plot, is a weak effect.  
We therefore seek a statistical measure using the entire supercluster sample. 

For each supercluster, we measure its inertia momentum tensor, 
${\bf I}^{s} \equiv (I^{s}_{ij})$, as
\begin{equation}
I^{s}_{ij} = \frac{1}{M_{s}}\sum_{\alpha}M^{\alpha}_{c}x^{\alpha}_{c, i}
x^{\alpha}_{c,j},
\label{eqn:ine}
\end{equation}
where $M^{\alpha}_{c}$ and ${\bf x}^{\alpha}_{c} \equiv (x^{\alpha}_{c,i})$ 
represent the mass and the position of the $\alpha$-th member cluster, 
respectively and $M_{s}$ is the total mass of the host supercluster. 
Then, we compute the eigenvalues and eigenvectors through the diagonalization 
of ${\bf I}^{s}$ and determine the major-axis direction as the direction of 
the eigenvector corresponding to the largest eigenvalue. 

It is, however, worth mentioning here that for a supercluster which has  
less than five clusters, the orientation of its major axis derived 
using equation (\ref{eqn:ine}) must suffer from considerable inaccuracy. 
The most idealistic way should be to derive the major axis of a supercluster 
using all particles within it.

Nevertheless, given that the major axes of the superclusters in 
real observations cannot be determined by this idealistic way 
since the positions of the dark matter particles are not measurable, the 
advantage of our analysis based (eq.[\ref{eqn:ine}]) is its practicality. 
That is, it can be readily repeated by observers based directly on cluster 
catalogs. 
 
At any rate, to overcome the limitation of our analysis based on equation 
(\ref{eqn:ine}, we construct a separate sample choosing only those  
superclusters which have more than five clusters ($N_{c} > 5$ where $N_{c}$ 
is the number of clusters within the superclusters).  
It is found that $217$ superclusters have more than five clusters 
and total $1492$ clusters belong to those $217$ superclusters.

\begin{figure}
\begin{center}
\plotone{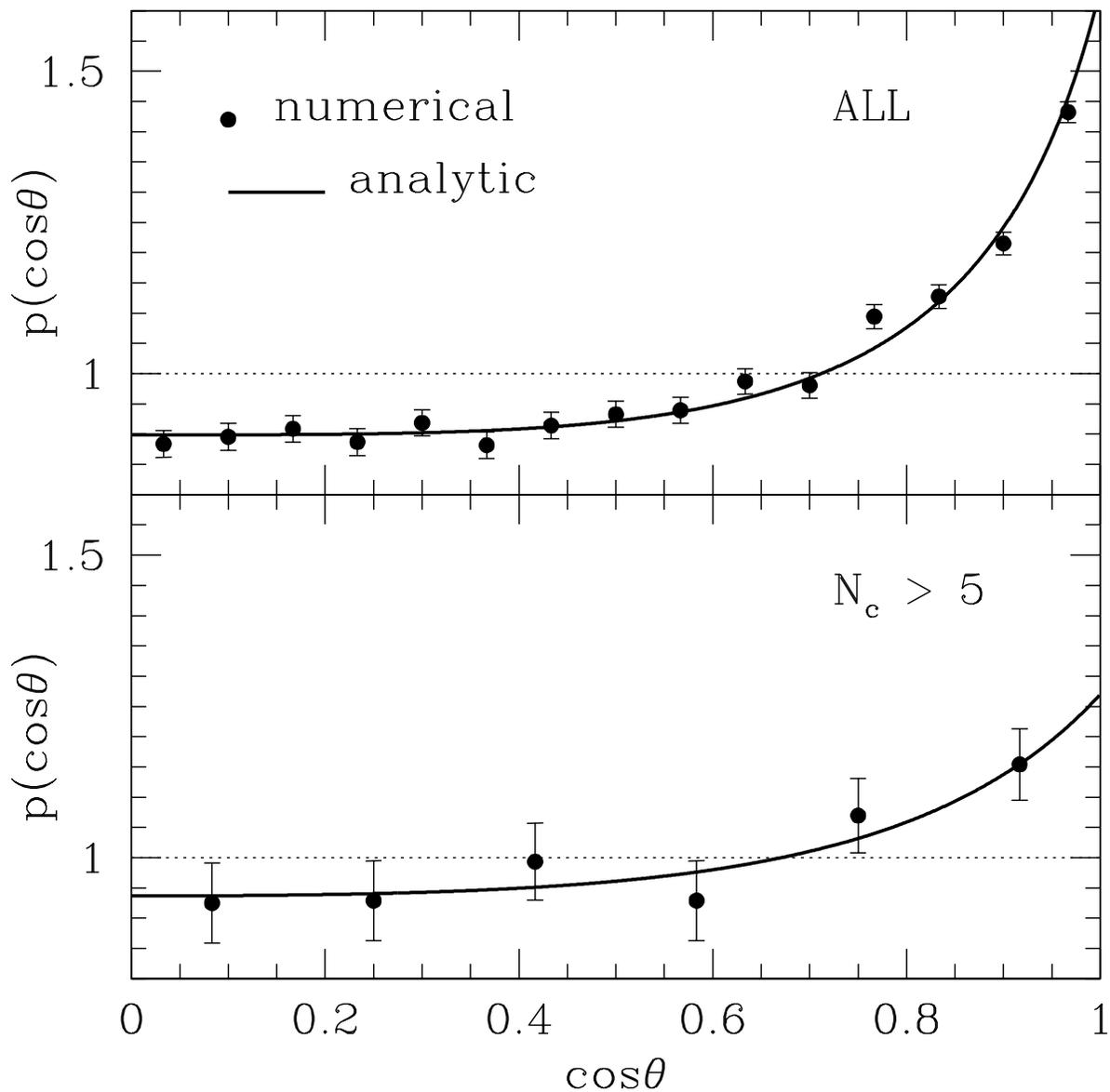}
\caption{Probability density distributions of the cosines of the
angles between the major axes of clusters and their superclusters: 
({\it Upper}): the case that all $14007$ superclusters are used; 
({\it Lower}): the case that only $217$ superclusters with more than 
five clusters are used. In each panel, the numerical result is represented 
by dots with Poissonian errors while the analytic result (\ref{eqn:theta_dis}) 
corresponds to the solid curve. The dotted line corresponds to the case of 
no alignment.
\label{fig:dis}}
\end{center}
\end{figure}

First, we measure the probability distribution of the cosines of the angles, 
$\cos\theta$, between the major axes of the superclusters and their member 
clusters. Figure \ref{fig:dis} plots the result as solid dots with Poisson 
errors. The upper panel corresponds to the case that all the $14,007$ 
superclusters are used, while the lower panel corresponds to the case 
that only those superclusters with more than five clusters are used. 
The dotted line in each panel corresponds to the case of no alignment. 
As can be seen, the distribution, $p(\cos\theta)$,  increases with 
$\cos\theta$ in the both panels, revealing a clear signal of alignment 
effect. Although the result of the lower panel shows less sharp increase, 
suffering from large errors, the signal is robust at the $99\%$ confidence 
level. It indicates that the cluster-supercluster alignment effect 
is not a false signal originated from the inaccurate derivation of 
the supercluster major axes but a real one. The mean value of $\cos\theta$ 
is found to be $0.54$ and $0.52$ in the upper and the lower panels, 
respectively.

\begin{figure}
\begin{center}
\plotone{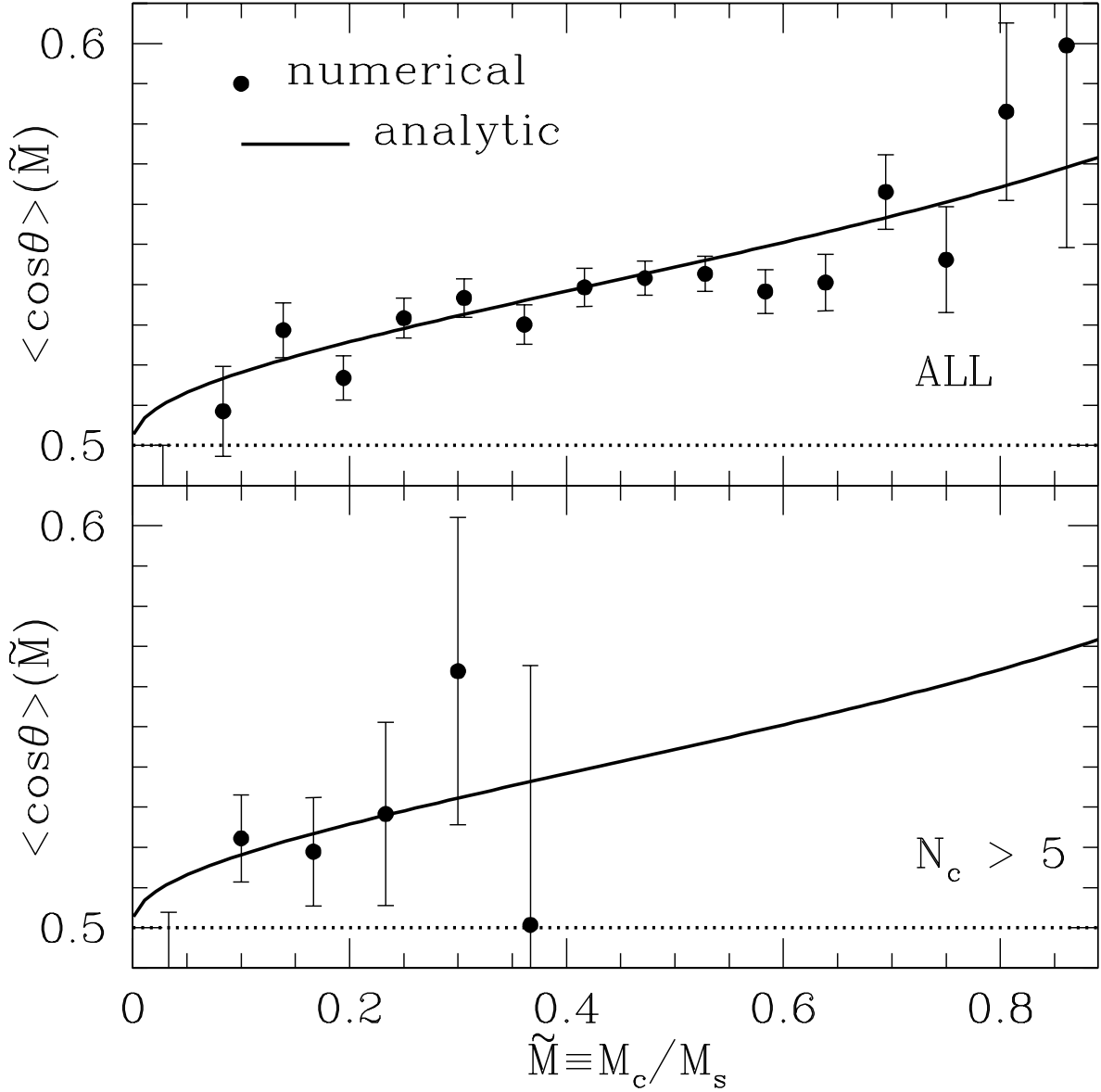}
\caption{Average of the cosines of the angles as a function of the
cluster mass: ({\it Upper}): the case that all $14007$ superclusters 
are used; ({\it Lower}): the case that only $217$ superclusters with 
more than five clusters are used. In each panel, the dots and solid curves 
represent the numerical and the analytic results, respectively. The errors 
are calculated as one standard deviation of the cosines of the angles for 
the case of no alignment. The dotted line corresponds to the case of no 
alignment.
\label{fig:mas}}
\end{center}
\end{figure}

Now that a robust signal of cluster-supercluster alignment effect is found,
we examine how the degree of the alignment depends on the cluster properties. 
First, we examine how the average of $\cos\theta$ depends on the cluster 
mass, $M_{c}$. Figure \ref{fig:mas} plots the result versus the rescaled 
cluster mass, $\tilde{M}\equiv M_{c}/M_{s}$,  as solid dots with errors 
which are calculated as one standard deviation of $\cos\theta$ for the case 
of no alignment. As can be seen in the upper panel, the degree of alignment 
increases with $\tilde{M}$. A similar trend is also shown in the lower panel 
although it suffers from the large errors.

\begin{figure}
\begin{center}
\plotone{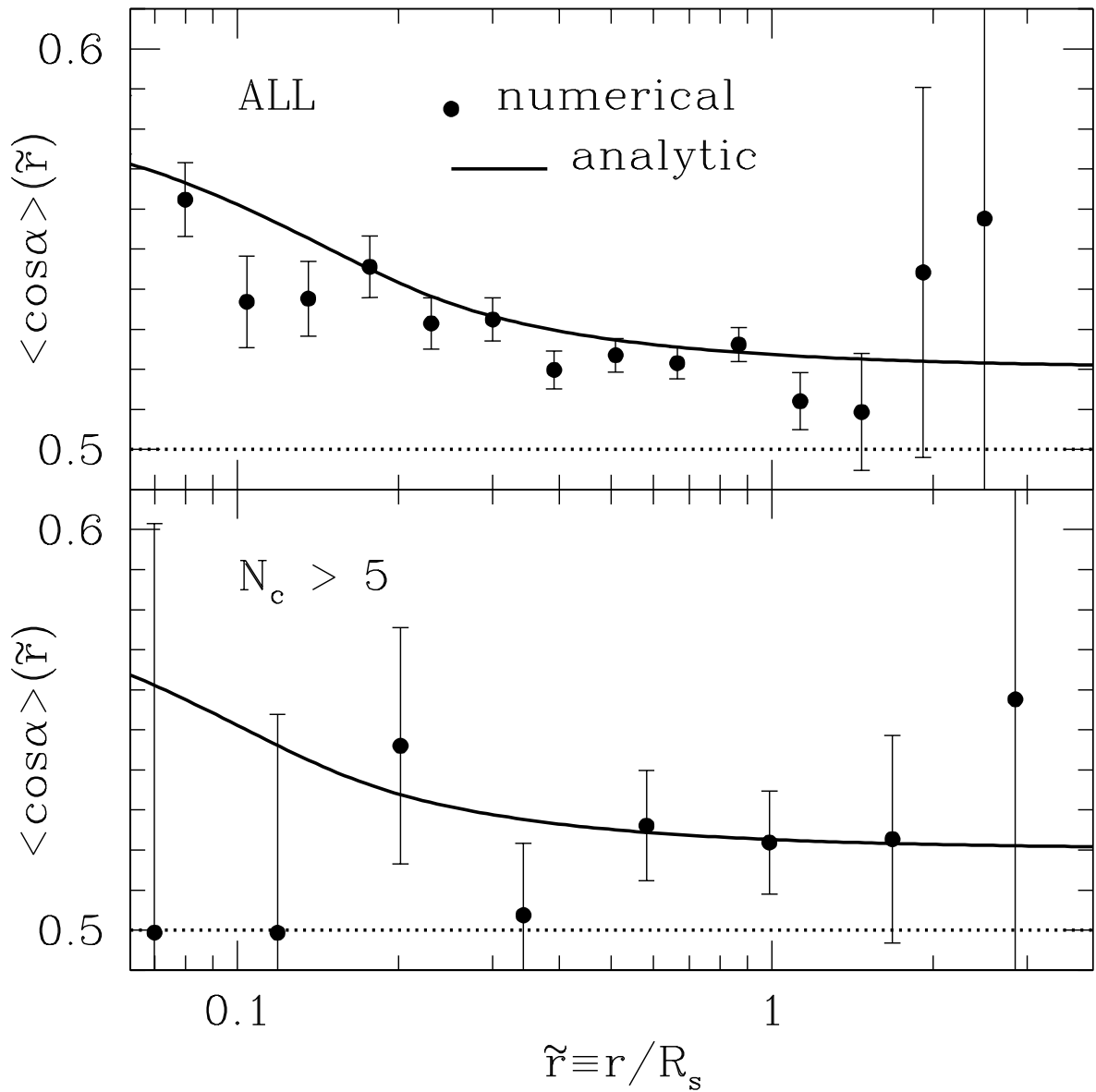}
\caption{Average of the cosines of the angles as a function of the
distance from the supercluster center to the cluster center: 
({\it Upper}): the case that all $14007$ superclusters are used; 
({\it Lower}): the case that only $217$ superclusters with more than 
five clusters are used. In each panel the dots and solid curves represent 
the numerical and the analytic results, respectively. The dotted line 
corresponds to the case of no alignment.
\label{fig:sep}}
\end{center}
\end{figure}

Second, we examine how the average of $\cos\theta$ depends on the separation 
distance, $r$, from the supercluster center to the cluster center. 
Figure \ref{fig:sep} plots the result versus the rescaled distance, 
$\tilde{r}\equiv r/R_{s}$ as solid dots with errors. 
As can be seen in the upper panel, the degree of the alignment decreases 
with distance. That is, the closer a cluster is located to the supercluster 
center,the more stronger the alignment effect is. 
 
\begin{figure}
\begin{center}
\plotone{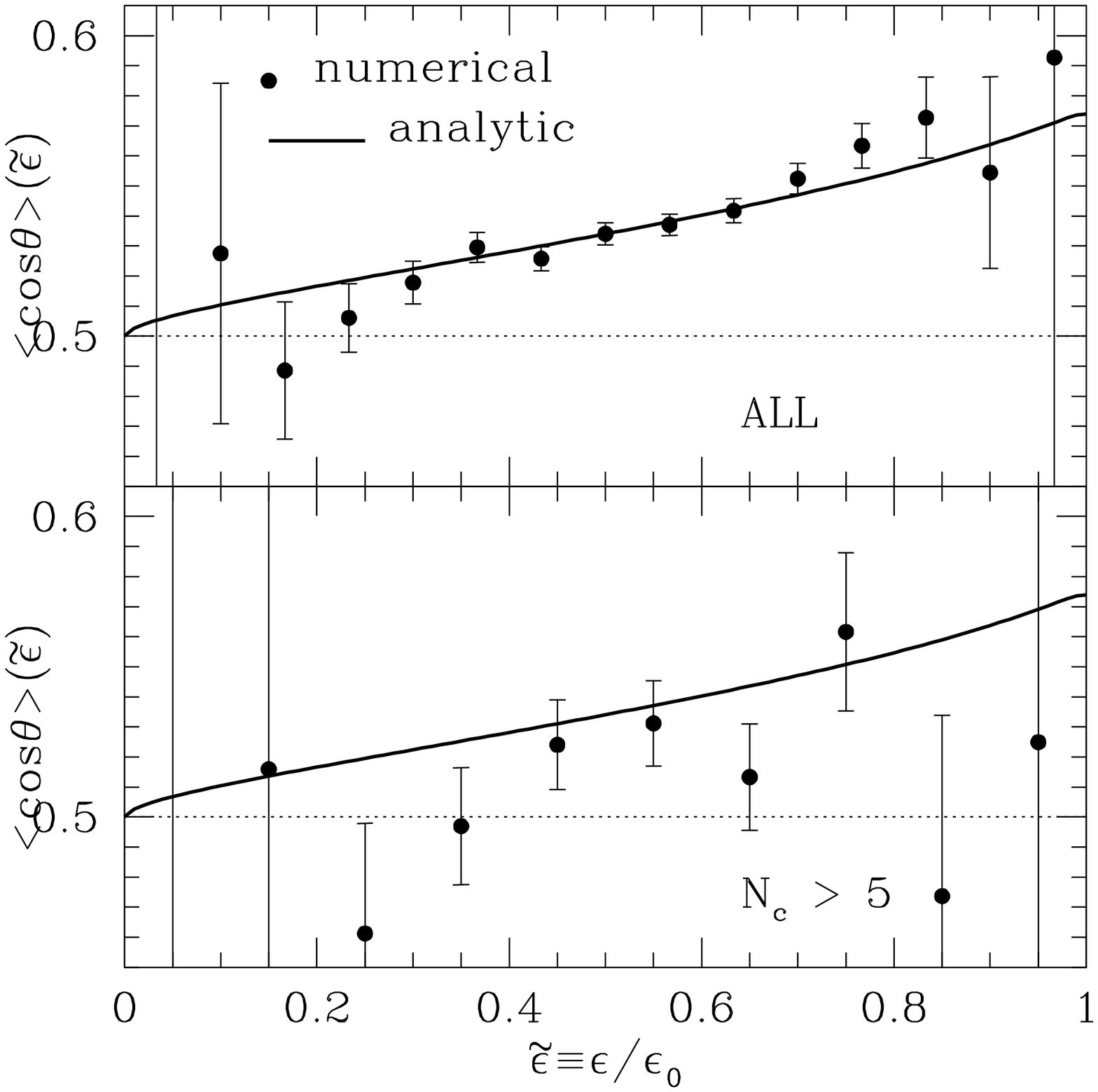}
\caption{Average of the cosines of the angles as a function of the
cluster ellipticity: ({\it Upper}): the case that all $14007$ 
superclusters are used; ({\it Lower}): the case that only $217$ superclusters 
with more than five clusters are used. In each panel, the dots and solid 
curves represent the numerical and the analytic results, respectively. 
The dotted line corresponds to the case of no alignment. 
\label{fig:ell}}
\end{center}
\end{figure}

Third, we examine how the average of  $\cos\theta$ depends on the  
cluster ellipticity, $\epsilon$. Here, we define the ellipticity of
a cluster as $\epsilon \equiv 1 - \sqrt{\varrho^{c}_{3}/\varrho^{c}_{1}}$ 
assuming a prolate cluster shape, where $\varrho^{c}_{1}$ and
$\varrho^{c}_{3}$ are the largest and the smallest eigenvalues of the
cluster inertia momentum tensor, respectively.  Figure \ref{fig:ell}
plots the result versus the rescaled ellipticity, $\tilde{\epsilon}\equiv 
\epsilon/\epsilon_{0}$ (where $\epsilon_{0}$ is 
the maximum cluster ellipticity) as solid dots with errors, which reveals 
that the degree of alignment increases with cluster ellipticity. 
That is, the more elongated a cluster is, the more stronger the 
alignment effect is.

\begin{figure}
\begin{center}
\plotone{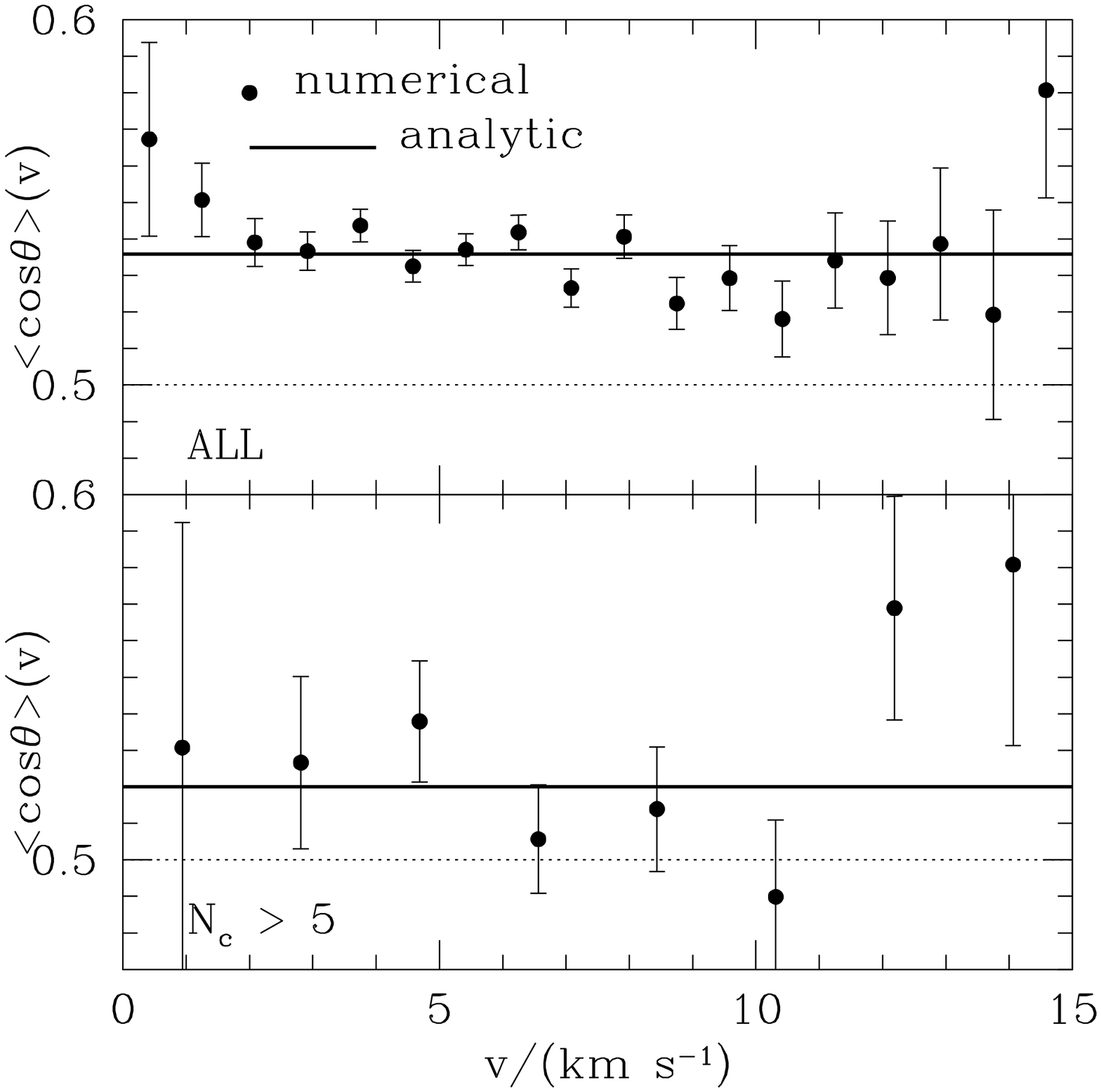}
\caption{Average of the cosines of the angles as a function of the
cluster velocity: ({\it Upper}): the case that all $14007$ superclusters 
are used; ({\it Lower}): the case that only $217$ superclusters with more than 
five clusters are used. In each panel, the dots and solid curves represent the
numerical and the analytic results, respectively. The dotted line corresponds 
to the case of no alignment.
\label{fig:vel}}
\end{center}
\end{figure}

Fourth, we measure the average of  $\cos\theta$ as a function of the
cluster velocity, $v$. Figure \ref{fig:vel} plots the result with solid
dots with errors, which reveals that the degree of alignment depends
very weakly on the cluster velocity.

We provide physical explanations for these numerical results in $\S 3$.
 
\section{PHYSICAL ANALYSIS}

\subsection{\it Hypotheses}
To construct an analytic model for the cluster-supercluster
alignment effect, we assume the following.
\begin{enumerate}
\item
A supercluster forms through anisotropic merging of clusters along
filaments. In consequence, the major axis of a supercluster tends to
be in the direction of the dominant filament. A filament is defined as
one dimensional object collapsed along the major and intermediate
principal axes of the local tidal tensor \citep{zel70,pog-etal98}. 
The direction of a filament thereby is aligned with the minor principal 
axis of the tidal tensor.  Therefore, the major axis of a supercluster 
tends to be in the direction of the minor principal axis of the tidal tensor. 
\item
Let ${\bf T}^{s}$ be the tidal tensor field smoothed on the supercluster mass 
scale, and let also $\delta_{s}\equiv\Delta\rho/\bar{\rho}$ be the linear 
density contrast of the supercluster where $\bar{\rho}$ is the mean mass
density. Let also $\lambda_1, \lambda_2,\lambda_3$ (with
$\lambda_1>\lambda_2 >\lambda_3$) be the three eigenvalues of ${\bf T}^{s}$.
The collapse condition for a supercluster is given as 
\begin{equation}
\label{eqn:sup}
\delta_{s} = \lambda_{1} + \lambda_{2} + \lambda_{3} =1.3,\qquad 
\lambda_1 > \lambda_2 > 0, \quad \lambda_3 < 0.
\end{equation}
Given that the supercluster passes the moment of turn-around but not
yet virialized, we expect its linear density contrast, $\delta_{s}$,
to be in the range $(1,1.68)$ where the values of $1$ and $1.68$
correspond to the linear densities at the moments of the turn-around
and the virialization, respectively \citep{eke-etal96}. Here, we
choose a fiducial value of $\delta_{s}=1.3$.  The other condition,
$\lambda_1> \lambda_2 > 0, \lambda_3 < 0$ in equation (\ref{eqn:sup})
represents the collapse along filaments \citep{pog-etal98}. 
\item
The cluster-supercluster alignment is a reflection of the anisotropic
spatial distribution of cluster galaxies in a filament-dominant
web-like cosmic structure. The correlation of the spatial positions of
galaxies with the local tidal field can be quantified by the following
quadratic equation which was first suggested by \citet{lee-kan06}.
\begin{equation}
\label{eqn:corr}
\langle {x}^{c}_{i}{x}^{c}_{j}|\hat{\bf T}^{s}\rangle = 
\frac{1 - s}{3}\delta_{ij} + s\hat{T}^{s}_{ik}\hat{T}^{s}_{kj},
\end{equation}
where ${\bf x}^{c}\equiv (x^{c}_{i})$ and $\hat{\bf T}^{s}=
(\hat{T}^{s}_{kj})\equiv T^{s}_{ik}/\vert{\bf T}^{s}\vert$ are the 
rescaled major axis of a galaxy cluster and the unit tidal shear
tensor smoothed on the supercluster mass scale, $M_{s}$. Here, the
parameter, $s \in [-1,1]$, represents the strength of the correlation
between ${\bf x}^{c}$ and  ${\bf T}^{s}$. If $s=-1$, there is the
strongest correlation between ${\bf x}^{c}$ and $\hat{\bf T}^{s}$. If
$s=1$, there is the strongest anti-correlation between ${\bf x}^{c}$
and $\hat{\bf T}^{s}$. While if $s=0$, there is no correlation between them.
\item
The conditional probability distribution of ${\bf x}^{c}$ provided
that the local tidal field is given as ${\bf T}^{s}$ can be
approximated as Gaussian \citep{lee-kan06}:
\begin{equation}
\label{eqn:xdis}
P({\bf x}^{c}|\hat{\bf T}^{s}) = \frac{1}{[(2\pi)^3 {\rm det}(M)]^{1/2}}
\exp\left[-\frac{x^{c}_{i}(M^{-1})_{ij}x^{c}_{j}}{2}\right],
\end{equation}
where the covariance matrix 
$M_{ij} \equiv \langle x^{c}_{i}x^{c}_{j}|\hat{\bf T}^{s}\rangle$ is
related to $\hat{\bf T}^{s}$ by equation (\ref{eqn:corr}). 
\end{enumerate}

It is worth mentioning here the difference of the cluster-supercluster
alignment from the galaxy-cluster alignment. For the former case, the
primordial tidal field induces the anisotropy in the spatial
distribution of galaxies along cosmic filaments, which results in the
alignment between the major axes of clusters and their superclusters.
For the latter case, the tidal field of a virialized cluster halo
induces the angular momentum of the cluster galaxies whose minor axes
tend to be aligned with the major axis of its host cluster
\citep{lee-etal05}. In other words, the alignments between the major
axes of cluster galaxies and their host clusters are related to the
generation of the angular momentum while the alignments between the
major axes of clusters and their host superclusters related to the 
filamentary distribution of galaxies. 

\subsection{\it Analytic Expressions}

Using the four hypotheses given in $\S 3.1$, we derive first $p(\cos\theta)$
analytically. According to the second hypothesis, it amounts to deriving the 
probability density distribution of the cosines of the angles between the 
major axes of clusters and the minor principal axes of the local tidal tensors.

Let us express ${\bf x}^{c}$ in terms of the spherical polar coordinates 
in the principal axis frame of ${\bf T}^{s}$ as ${\bf x}^{c} = 
(x^{c}\sin\theta\cos\phi,x^{c}\sin\theta\sin\phi,x^{c}\cos\theta)$ 
where $x^{c}\equiv \vert{\bf x}^{c}\vert$ and $\theta$ and $\phi$ are the
polar and the azimuthal angles of ${\bf x}^{c}$, respectively. Then, 
the polar angle, $\theta$, is nothing but the angle between ${\bf x}^{c}$ 
and the minor principal axis of ${\bf T}^{s}$. Now, the probability 
density distribution of $\cos\theta$ can be derived by integrating 
equation (\ref{eqn:xdis}) over $x^{c}$ and $\phi$ as  $p(\cos\theta) = 
\int_{0}^{2\pi}\int_{0}^{\infty}P(x^{c},\theta,\phi)x^{c 2}dx^{c}d\phi$, 
which leads to \citep{lee-kan06}
\begin{eqnarray}
\label{eqn:theta_dis}
p(\cos\theta) &=& \frac{1}{2\pi}\prod_{i=1}^{3}
\left(1-s+3s\hat{\lambda}^{2}_{i}\right)^{-\frac{1}{2}}\times \nonumber \\
&&\int_{0}^{2\pi}
\left(\frac{\sin^{2}\theta\cos^{2}\phi}{1-s+3s\hat{\lambda}^{2}_{1}} + 
\frac{\sin^{2}\theta\sin^{2}\phi}{1-s+3s\hat{\lambda}^{2}_{2}} + 
\frac{\cos^{2}\theta}{1-s+3s\hat{\lambda}^{2}_{3}}\right)^{-\frac{3}{2}}
d\phi, 
\end{eqnarray}
where $\{\hat{\lambda_i}\}_{i=1}^{3}$ are the unit eigenvalues of
$\hat{\bf T}$ related to $\{\lambda_i\}_{i=1}^{3}$ as 
$\hat{\lambda_i} \equiv \lambda_{i}/
\left(\lambda^{2}_{1}+\lambda^{2}_{2}+\lambda^{2}_{3}\right)^{1/2}$. 

It was in fact \citet{lee-kan06} who first derived equation
(\ref{eqn:theta_dis}) as an analytic expression for the probability
distribution of the alignments between the positions of the galaxy
satellites in the major axis orientations of their host
galaxies. Here, we derive it as an analytic expression for the
probability distribution of the alignments between the major axes of
clusters and their host superclusters.  
It is important to note a key difference between the two cases. For
the case of galaxy satellites, it is the tidal fields of the {\it
virialized} galactic  halos that causes the alignment
effect. Therefore, all the eigenvalues, $\lambda_{1}, \lambda_{2},
\lambda_{3}$ in equation (\ref{eqn:theta_dis}) should be positive.
Whereas for the case of  clusters in superclusters it is the local
filaments which collapse along only two principal axes of the
primordial local tidal tensors. Therefore, $\lambda_{3}$ in equation
(\ref{eqn:theta_dis}) has a negative value.

The probability distribution, $p(\cos\theta)$
(eq.[\ref{eqn:theta_dis}]) is characterized by three independent
parameters,  $s$, $\lambda_1$, and $\lambda_2$. Once the values of
$\lambda_{1}$ and $\lambda_{2}$ are determined, then the negative value of
$\lambda_{3}$ is automatically determined by equation (\ref{eqn:sup}). 
Since the values of these three parameters depend on the properties of
individual superclusters as well as the local conditions of the
initial tidal fields, it may be quite difficult to determine them
analytically.
  
Instead, we determine their average values by fitting equation
(\ref{eqn:theta_dis}) to the numerical results obtained in $\S 2$.
When the numerical result using all superclusters are fitted, 
the best-fit values of the three parameters are found to be $s = -0.71$, 
$\lambda_{1}=2.23$, and $\lambda_{2}=0.53$, which gives $\lambda_3=-1.46$.  
When the numerical result using only those superclusters with $N_{c}>5$ are 
fitted, it is found interestingly that the best-fit values of the parameters 
are $s=-0.5$, $\lambda_{1}=2.23$, and $\lambda_{2}=0.53$. Note that the 
two numerical cases yield the same best-fit values for $\lambda_{1}$ and 
$\lambda_{2}$ although the best-fit values of the correlation parameters 
are different as $s=-0.71$ and $s=-0.5$.

Figure \ref{fig:dis} plots the analytic distributions with these best-fit 
parameters (solid line) and compares it with the numerical data points. 
In the upper panel, the analytic distribution with $s=-0.71$ is compared 
with the numerical result obtained in $\S 2$ using all superclusters,  
while in the lower panel the analytic distribution with $s=-0.5$ is compared 
with the numerical result obtained using only those superclusters with more 
than five clusters $(N_{c} > 5$). As can be seen, the analytic and the 
numerical results are in good agreement with each other in the both panels.

It is worth mentioning here that the best-fit values of the three
parameters are subject to our fiducial choice of $\delta_{s}=1.3$. As
mentioned in $\S 3.1$, there is no consensus on the critical linear
density of the superclusters unlike the case of clusters. Varying the
value of $\delta_{s}$ from $1.0$ to $1.68$, we have repeated the
fitting procedure, and we found that although the best-fit values of
$\lambda_{1}$ and $\lambda_{2}$ change by maximum $20\%$, the fitting 
result itself does not sensitively change with the value of $\delta_{s}$. 
Thus, it is concluded that our fiducial model is a stable choice.

Now that we have the probability density distribution, $p(\cos\theta)$, 
we would like to find an analytic expression for $\langle\cos\theta\rangle$ 
as a function of cluster mass, position, ellipticity, and velocity. 
The dependence of the correlation parameter $s$ on the cluster mass $M$ 
may be obtained by considering the difference in mass between the cluster 
and its host supercluster. Strictly speaking, equation (\ref{eqn:corr}) is 
valid when the tidal tensor ${\bf T}^{s}$ and the position vector 
${\bf x}^{c}$ are smoothed on the same mass scale. In other words, the 
correlation between ${\bf x}^{c}$ and ${\bf T}^{s}$ is expected to be highest 
when the two smoothing mass scales are the same. In reality, however, 
${\bf T}^{s}$ is smoothed on the supercluster mass scale $M_{s}$ while 
${\bf x}^{c}$ is smoothed on the cluster mass scale $M_{c}$. The difference 
between the two mass scales diminishes the correlation between ${\bf x}^{c}$ 
and ${\bf T}^{s}$.  

Let $s_{M_{0}}$ be the value of the correlation parameter when the 
tidal field is smoothed on the same cluster mass scale, ${\bf T}^{c}$. 
We expect $s_{M_0}=-1$. Given equation (\ref{eqn:corr}), we approximate 
$s=s(\tilde{M})$ as 
\begin{equation}
\label{eqn:sm}
s(\tilde{M}) \approx s_{M_0}\frac{\langle\hat{T}^{s}_{ik}
\hat{T}^{s}_{kj}\rangle}
{\langle\hat{T}^{c}_{ik}\hat{T}^{c}_{kj}\rangle}
\approx s_{M_0}\frac{\sigma^{2}_s}{\sigma^{2}_c}.
\end{equation}
Here, $\sigma_c$ and $\sigma_s$ represent the rms linear density fluctuations 
smoothed on the mass scales of $M_c$ and $M_s$, respectively. In deriving 
equation(\ref{eqn:sm}) we use the approximation of 
$\langle\hat{T}^{c}_{ik}\hat{T}^{c}_{kj}\rangle \approx 
\langle {T}^{c}_{ik}{T}^{c}_{kj}\rangle/\vert{\bf T}^{c}\vert^{2}$, 
which was proved to be valid by \citet{lee-pen01}. 

Now that the functional form of $s(\tilde{M})$ is found, the average of
$\cos\theta$ as a function of $\tilde{M}$ can be calculated by equations
(\ref{eqn:theta_dis}) and (\ref{eqn:sm}) as 
\begin{equation}
\label{eqn:mas}
\langle\cos\theta\rangle(\tilde{M})=\int^{\infty}_{0}\cos\theta
p[\cos\theta;s(\tilde{M})]d\!\cos\theta.
\end{equation}
Figure \ref{fig:mas} plots equation (\ref{eqn:mas}) with $\lambda_{1}=2.23$ 
and $\lambda_{2}=0.53$ (solid line), and compares it with the numerical 
result (dots) obtained in $\S 2$. For the analytic distribution, the value 
of $M_{s}$ is set to be the mean mass of the superclusters found in $\S 2$: 
$1.26 \times 10^{15}h^{-1}M_{\odot}$ (upper panel); 
$3.69 \times 10^{15}h^{-1}M_{\odot}$ (lower panel).
As can be seen, in the upper panel the analytic and the numerical results 
agree with each other excellently. In the lower panel, although the numerical 
result suffers from large errors, the analytic prediction is still quite 
consistent with the numerical result. 

The dependence of the correlation parameter, $s$, on the distance, $r$,
between the centers of clusters and their host superclusters can be
obtained in a similar way. The correlation between ${\bf x}^{c}$ and
${\bf T}^{s}$ in equation (\ref{eqn:corr}) becomes strongest when
$r=0$. In reality, however, $r$ always deviates from zero, which will
diminish  the correlation strength.

Let $s_{r0}$ be the value of the correlation parameter when $r=0$,
which is expected again  to be $s_{r0} = -1$. With a similar
approximation made for equation (\ref{eqn:sm}), we find the following formula 
for $s(r)$: 
\begin{equation}
\label{eqn:sr}
s(r) \approx \frac{1}{2}s_{r0}\left[1 + 
\frac{\langle\hat{T}^{c}_{ij}({\bf x}+{\bf r}){\hat T}^{c}_{ij}
({\bf x})\rangle}
{\langle\hat{T}^{c}_{ij}({\bf x})\hat{T}^{c}_{ij}({\bf x})\rangle}\right]
\approx \frac{1}{2}s_{r0}\left[1 + \tilde{\xi}_{c}(r)\right],
\end{equation}
where $\tilde{\xi}_{c}(r)$ represents the two point density correlation
rescaled to satisfy $\tilde{\xi}_{c}(0)=1$. Since the distance $r$ is
a Eulerian quantity unlike the mass $M_{s}$, we use the Eulerian
filtering  radius of $2h^{-1}$Mpc, the typical cluster size,  to
convolve the correlation function $\tilde{\xi}_{c}$. Here, the factor
of $1/2$ comes from the average decreases of the correlation parameter
due to the mass difference between the clusters and their host
superclusters. 

Now that the functional form $s(r)$ is found, the average of
$\cos\theta$ as a function of $r$ can be calculated through
equations (\ref{eqn:theta_dis}) and (\ref{eqn:sr}) as 
\begin{equation}
\label{eqn:sep}
\langle\cos\theta\rangle(r)=\int^{\infty}_{0}\cos\theta
p[\cos\theta;s(r)]d\!\cos\theta.
\end{equation}
Figure \ref{fig:sep} plots equation (\ref{eqn:sep}) (solid line) 
as a function of the rescaled distance, $\tilde{r}\equiv r/R_{s}$, and 
compares it with the numerical result (dots) obtained in $\S 2$. 
For the analytic distribution, the value of $R_{s}$ is set to be the mean 
Lagrangian radius of superclusters found in $\S 2$ using the relation 
of $R_{s} = [3\bar{M}_{s}/(4\pi\bar{\rho})]^{1/3}$. As can be seen, 
the analytic and the numerical results agree with each other 
quite well. 

Regarding the dependence of $s$ on the cluster ellipticity,
$\epsilon$, although it is predicted qualitatively in our theoretical
model that the degree of the alignment increases with ellipticity, the
quantitative functional form of $s(\epsilon)$ is quite difficult to
determine analytically since the cluster ellipticity are sensitively
vulnerable to modifications caused by nonlinear merging and infall process. 

Instead of using analytic approach, numerical fitting is used to
determine the functional form of $s(\epsilon)$. Let $s_{\epsilon 0}$
represents the value of $s$ when the cluster ellipticity has the maximum 
value, $\epsilon 0$. It is expected again that $s_{\epsilon 0}=-1$. 
We find that the following formula gives a good fit to the numerical results
\begin{equation}
\label{eqn:se}
s(\tilde{\epsilon}) =  s_{\epsilon 0}\tilde{\epsilon}^{1/2}. 
\end{equation}
Since $\epsilon_{0}$ is defined as the maximum ellipticity, 
$s(\tilde{\epsilon})$ has the extreme value of $-1$ at $\tilde{\epsilon}=1$. 
Note that the value of $\epsilon_{0}$ is not fixed but sample-dependent.  
Here, the Millennium Run data we use yields $\epsilon_{0}=0.7$. 
But, a different sample could yield a different value of $\epsilon_{0}$. 

Now that the functional form of $s(\tilde{\epsilon})$ is found, the average 
of $\cos\theta$ as a function of $\tilde{\epsilon}$ can be calculated
by equations (\ref{eqn:theta_dis}) and (\ref{eqn:se}) as 
\begin{equation}
\label{eqn:ell}
\langle\cos\theta\rangle(\tilde{\epsilon})=\int^{\infty}_{0}
\cos\theta p[\cos\theta;s(\tilde{\epsilon})]d\!\cos\theta. 
\end{equation}
The comparison between the analytic result (eq.[\ref{eqn:ell}]) and the 
numerical data points shows good consistency, as seen in Fig. \ref{fig:ell}. 

Regarding the dependence of $s$ on the cluster velocity $v$, no
strong dependence is expected in our model since the primordial tidal
field is uncorrelated with the velocity field \citep{bar-etal86}. 
Therefore, we model it as an uniform distribution as 
$\langle\cos\theta\rangle(v) =\langle\cos\theta\rangle$. The average 
value, $\langle\cos\theta\rangle$ is found to be $0.54$ when $s=-0.71$ 
(upper panel) while it is $0.52$ when $s=-0.5$ (lower panel).  
As can be seen in Fig. \ref{fig:vel},  the analytic and the numerical 
results are consistent with each other.

\section{DISCUSSION AND CONCLUSION}

In the context of the standard cosmic web picture of large-scale
structure, we have constructed a parametric model for the alignment of
cluster-sized halos with their host superclusters.  
The underlying assumption is that cluster-supercluster alignment reflects 
the spatial distribution of matter as it is organized along
filamentary structures by the primordial tidal field.  
The parameters of the analytic model represent the dominance of
filaments and the spatial coherence of the initial tidal field.

We show that the analytic model provides a good fit to orientation data 
derived from mass-limited halo samples of a $\Lambda$CDM Hubble volume
simulation.  After fitting the three free parameters using the overall 
distribution of cluster-supercluster alignment angles, the model then 
simultaneously matches the behavior of the mean alignments as a
function of relative mass, cluster position within the supercluster,
and cluster ellipticity.  No trend with cluster velocity is predicted
or measured in the simulation. 

It is worth discussing a couple of simplified assumptions on which our
theoretical model is based. First,we have used the FOF algorithm to
identify superclusters in the N-body simulation data. Unlike the case
of virialized clusters, however, there is no established consensus on
how to define superclusters. Different supercluster-identification
algorithms could result in different multiplicity, mass, and shape of
superclusters which would in turn affect our results. 

Second, we have assumed that the filaments correspond to the
Lagrangian regions where only the largest and the second largest
eigenvalues of the local tidal tensor are positive. Although this
definition of a filament is consistent with the picture of the
Zel'dovich approximation, it is obviously an oversimplification of the
reality. A more realistic definition and treatment of cosmic filaments
will be necessary to refine the model.

Another issue that we would like to discuss here is the possibility of
using the cluster-supercluster alignment effect as a cosmological
probe. We have shown that the phenomena of cluster-supercluster
alignments are closely related to the dominance of filaments, the
web-like distribution of galaxies on very large scales.
The dominance of filaments is in turn related to the spatial
correlations of the primordial tidal field, which depends sensitively on
the slope of the initial power spectrum on the supercluster
scale. Thus, by measuring the degree of cluster-supercluster alignment,
it might be possible to constrain the slope of the initial power
spectrum in a complementary way. 

Finally, we conclude that our model for the cluster-supercluster
alignments will provide a theoretical framework within which the
distribution of cosmic structures on the largest scales can be
physically understood and quantitatively described.

\acknowledgments
We are grateful to the anonymous referee who helped us improve the original 
manuscript. We are also grateful to the warm hospitality of Y.Suto and the 
University of Tokyo, where this work was initiated.  J.L. also thanks D. Park 
for useful helps. J.L. is supported by the research grant No. 
R01-2005-000-10610-0 from the Basic Research Program of the Korea Science and 
Engineering Foundation. A.E.E. acknowledges support from the Miller Institute 
for Basic Research in Science at the University of California, Berkeley, 
from NSF ITR ACI-0121671 and from NASA ATP NAG5-13378.

\clearpage

\end{document}